\documentclass{PoS}
\usepackage{epsfig}
\usepackage{cite}
\usepackage{enumitem}

\newcommand{\postscript}[2]{\setlength{\epsfxsize}{#2\hsize}
   \centerline{\epsfbox{#1}}}
\newcommand{\comment}[1]{}

\usepackage[usenames,dvipsnames]{xcolor}
\definecolor{orange}{cmyk}{0,0.5,1,0}
\definecolor{rossoCP3}{cmyk}{0,.88,.77,.40}
\definecolor{graa}{rgb}{0.8,0.8,0.8}
\definecolor{blaa}{rgb}{0.2,0.2,0.6}

\title{The Galactic
    magnetic field in the light of starburst-generated
    ultrahigh-energy cosmic rays}

\ShortTitle{GMF in the light of starburst-generated UHECRs}

\author{\speaker{Jorge F. Soriano} \\
Department of Physics \& Astronomy,  Lehman College, CUNY, NY 10468, USA\\
Department of Physics,
 Graduate Center, City University
  of New York,  NY 10016, USA\\
E-mail: \email{jfdezsoriano@gmail.com}}

\author{Luis A. Anchordoqui\\
Department of Physics \& Astronomy,  Lehman College, CUNY, NY 10468, USA\\
Department of Physics,
 Graduate Center, City University
  of New York,  NY 10016, USA\\
Department of Astrophysics,
 American Museum of Natural History, NY
 10024, USA\\
        E-mail: \email{luis.anchordoqui@gmail.com}}

\author{Diego F. Torres\\
Institute of Space Sciences (ICE-CSIC),  Campus UAB,
  Carrer de Magrans s/n, 08193 Barcelona, Spain \\
Instituci\'o Catalana de Recerca i Estudis Avan\c{c}ats
  (ICREA),  E-08010 Barcelona, Spain\\
Institut d'Estudis Espacials de Catalunya (IEEC),
08034 Barcelona, Spain\\
 E-mail: \email{dtorres@ice.csic.es}}

      \abstract{Auger data show evidence for a correlation between ultrahigh-energy
  cosmic rays (UHECRs) and nearby starburst galaxies. This intriguing
  correlation is consistent with data collected by the Telescope
  Array, which have revealed a much more pronounced directional ``hot
  spot'' in arrival directions not far from the starburst galaxy
  M82. In this work, we assume starbursts are sources of UHECRs and 
  investigate the prospects to use the observed distribution of UHECR
  arrival directions to constrain Galactic magnetic field models. We
  show that if the Telescope Array hot spot indeed originates from M82,
  UHECR data would place a strong constraint on the coherent and turbulent components
  of the Galactic magnetic field.
}

\FullConference{36th International Cosmic Ray Conference - ICRC219 -\\
		July 24 - August 1, 2019\\
	 Madison, Wisconsin, USA}

\begin{document}

\section{General Idea}

Magnetic fields are one of the most challenging astrophysical
phenomena to measure. We do have indications that magnetic fields are
everywhere in the Universe, but they are often very weak and
challenging to characterize in detail. The Milky Way is host to a
magnetic field on the order of $10^{-6}$~G, which is nearly a million
times smaller than the Earth's magnetic field. We know magnetic fields
exist on galaxy scales and larger in the Universe, but we do not know
how they got there, neither we do completely understand their role in
how the Universe has evolved.  Our observational and theoretical
understanding of magnetic fields in the Milky Way and of the global
structure of the Galactic magnetic field (GMF) has matured over many
decades~\cite{Beck:1995zs,Widrow:2002ud}, with a new-generation of more sophisticated and
quantitatively-constrained models emerging in the last
decade~\cite{Pshirkov:2011um,Jansson:2012pc,Jansson:2012rt,Beck:2014pma,Unger:2017wyw,Unger:2019xct,Han}.

GMF models are constrained by:
\begin{description}[noitemsep,topsep=0pt]
\item {\it (i)~Multi-frequency radio observations of the Faraday rotation
of extragalactic radio sources.} The polarization plane of a linearly polarized electromagnetic wave which propagates through a magnetized plasma rotates by an angle $\psi$ proportional to
the square of the wavelength $\lambda$, i.e. $\Delta \psi = {\rm RM}
\, \lambda^2$. To determine the rotation measure RM we then require multi- or at
least bi-frequency observations. The value of RM is proportional to
the line-of-sight integral
\begin{equation}
  {\rm RM} = c_1 \int_0^D dx_3 \ n_e \ B_\parallel \,,
\label{RM}
\end{equation}
where $c_1 \simeq 2.7 \times 10^{-23}~{\rm rad}/\mu {\rm G}$ is the
proportionality constant, $B_\parallel$ is the the longitudinal
component of the GMF, $D$ is the distance to the source, and and $n_e$
is the density of thermal electrons of the warm ionized medium of the
Galaxy~\cite{Pshirkov:2013wka}.
\item {\it (ii)~Measurements of the polarized synchrotron emission of
cosmic-ray electrons.} Galactic synchrotron emission sets a constraint
on the GMF that is complementary to the one from RMs, because synchrotron emission
depends on the transverse GMF $B_\perp$,
weighted by the relativistic (a.k.a. cosmic-ray) electron density
$n_{{\rm CR}e}$. The polarization state of linearly polarized light is specified by the
Stokes parameters Q and U, with each proportional to the polarized
intensity (PI) 
\begin{equation} {\rm PI}_i \sim \int_0^D  dx_3 \ \epsilon_{ij3} \
  n_{{\rm CR}e} \ B_j^2 \, ,
  \label{PI}
\end{equation}
where $B_i$ with $i =  (1,2)$ are the components of $B_\perp$, and
$\epsilon_{ij3}$ are components of the Levi-Civita tensor $\epsilon_{ijk}$~\cite{Farrar:2014hma}. The orientation of $B_\perp$ can be inferred from the Stokes parameters Q and U.
\end{description}
Exploiting the interconnections (\ref{RM}) and (\ref{PI}) between the
GMF and the physical observables depends on our understanding of the
thermal and relativistic electron distributions.

The GMF also deflects ultra-high energy cosmic rays (UHECRs). For a
cosmic ray of energy $E$ and charge $Ze$, the arrival direction
$\vec \xi$ is related to the the point of entry into the Galaxy
$\vec \zeta$ according to $\vec \zeta= \vec \xi + \vec \delta $, where
$\vec \delta$ is proportional to the line-of-sight integral
\begin{equation}
  \delta_i = c_2 \int_0^D  dx_3 \ \epsilon_{ij3} \ B_j^2 \, ,
  \label{delta}
\end{equation}
with $c_2= Ze/(E \, \mu{\rm G})$. The similarities between
(\ref{RM}), (\ref{PI}), and (\ref{delta}) suggest that knowing
the nuclear composition of UHECRs and each point of entry into the
Galaxy, the distribution of arrival directions provides a robust
constraint on the GMF.

The current upper limit on the extragalactic magnetic field is
$B \sim 1~{\rm nG}$~\cite{Pshirkov:2015tua}, and so the typical 
deflection from a source 3.5~Mpc away is estimated to be~\cite{Waxman:1996zn,Farrar:2012gm}
\begin{equation}
\delta \theta_{\rm eg} \lesssim 1.5^\circ Z \left(\frac{E}{10^{10}~{\rm
      GeV}}\right)^{-1} \, .
\end{equation}
For reasons outlined below, herein we are interested in UHECR nuclei of 
$Z  \leq 8$ and $E \gtrsim 10^{10.6}~{\rm GeV}$.
For nearby sources, the expected deflections of
these nuclei on the extragalactic magnetic field are $\lesssim 3^\circ$.  This implies that the galactic longitude $l$
and latitude $b$ indicating the UHECR point of entry into the
Galaxy are roughly coincident with the coordinates of the source
location.

In summary, to constrain the GMF using UHECR observations we need
high-resolution measurements of the arrival direction distribution and
the mass spectrum, and we also need to identify the sources. To
successfully fit our guides, we start out in the next section by
pinpointing a possible correlation between UHECRs and starburst
galaxies (SBGs), as a first step in the source identification. In the last
section, we estimate the expected deflections from nearby starbursts on
the basis of existing GMF models and comment on the prospects to
measure the mass spectrum with future experiments.

\section{Setting the Stage}

We have long been suspecting that SBGs are sources of
UHECRs~\cite{Anchordoqui:1999cu}. Over
the years, stronger and stronger experimental evidence has been
accumulating indicting a possible correlation between the arrival
directions of the highest energy cosmic rays and nearby SBGs~\cite{Anchordoqui:2002dj,Nemmen:2010bp,Aab:2017njo,Aab:2018chp}. 

Using data collected by the Pierre Auger Observatory,
the hypothesis of UHECR emission from the 23 brightest nearby
SBGs with a radio flux larger that 0.3~Jy
(selected out the 63 objects within 250~Mpc search for $\gamma$-ray
emission by the Fermi-LAT Collaboration~\cite{Ackermann:2012vca}) was
tested against the null hypothesis of isotropy through an unbinned
maximum-likelihood analysis~\cite{Aab:2018chp}. The adopted test
statistic (TS) for deviation from isotropy being the standard
likelihood ratio test between the starburst-generated UHECR sky model
and the null hypothesis.  The TS was maximized as a function of two
free parameters (the angular radius common to all sources, which
accounts in an effective way for the magnetic deflections, and the
signal fraction), with the energy threshold varying in the range
$10^{10.3} \lesssim E/{\rm GeV} \lesssim 10^{10.9}$. For a given energy
threshold, the TS for isotropy follows a $\chi^2$ distribution with
two degrees of freedom. The TS is maximum above $10^{10.6}~{\rm GeV}$,
with a local $p$-value of $3 \times 10^{-6}$. The smearing angle and
the anisotropic fraction corresponding to the best-fit parameters are
${13^{+4}_{-3}}^\circ$ and $(10 \pm 4)\%$, respectively. Remarkably,
the energy threshold of largest statistical significance coincides
with the observed suppression in the
spectrum~\cite{Aab:2017njo}, implying
that when we properly account for the barriers to UHECR propagation in
the form of energy loss
mechanisms~\cite{Greisen:1966jv,Zatsepin:1966jv} we obtain a self
consistent picture for the observed UHECR horizon. The scan in energy
thresholds comes out with a penalty factor, which was estimated
through Monte-Carlo simulations.  The post-trial chance probability in
an isotropic cosmic ray sky is $4.2 \times 10^{-5}$, corresponding to
a 1-sided Gaussian significance of $4\sigma$~\cite{Aab:2018chp}.

Very recently, the Telescope Array (TA) Collaboration carried out a
test of the reported correlation between the arrival directions of
UHECRs and SBGs~\cite{Aab:2018chp}. The data are compatible with
isotropy to within $1.1\sigma$ and with Auger result to within
$1.4\sigma$, and so the TA Collaboration concluded that with their
current statistics they cannot make a statistically significant
corroboration or refutation of the reported possible correlation
between UHECRs and SBGs~\cite{Abbasi:2018tqo}. However, TA data have
revealed a pronounced directional ``hot spot''~\cite{Abbasi:2014lda}
in arrival directions not far from the starburst galaxy
M82~\cite{He:2014mqa,Pfeffer:2015idq}. In this work we show that if
the TA hot spot indeed originates from M82, UHECR data would place a
strong constraint on  GMF models.

\begin{figure}
    \postscript{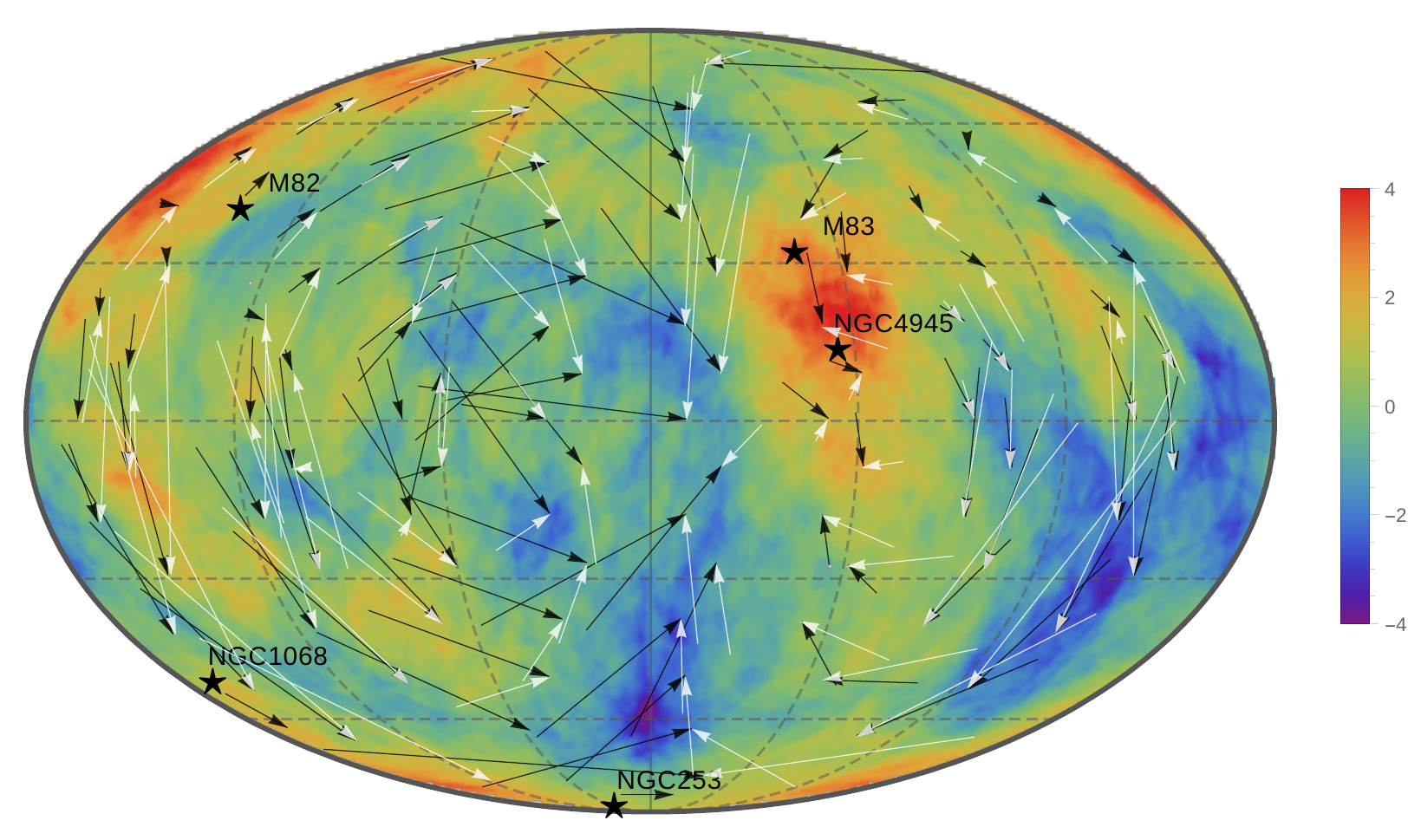}{0.73}
    \caption{Skymap in Galactic coordinates of the Li-Ma significances
      of overdensities in $20^\circ$ radius windows for 840 events
      recorded by Auger with $E > E_{\rm Auger}$ and 130 events
      recorded by TA with $E > E_{\rm TA }$. The color scale indicates
      the significance in units of standard deviations; negative
      values follow the convention of indicating the (positive)
      significance of deficits. We have superimposed the expected
      deflections from UHECR protons with $E = 10^{10}~{\rm GeV}$ as
      predicted by the PTKN (white)~\cite{Pshirkov:2011um} and JF
      (black)~\cite{Jansson:2012pc,Jansson:2012rt}
      models~\cite{Mollerach:2017idb,Farrar:2017lhm,Anjos:2018mgr}. The
      beginning of the arrows indicate the location of the sources and the
      tip of the arrows indicate the arrival direction on Earth. The
      Galactic Center is at the center of the skymap. The RGB color
      components of the skymap and legend presented
      in~\cite{Biteau:2019aaf} were sampled taking enough points per
      pixel to ensure that no information is loss. To each point
      sampled from the skymap, we associate a value for the Li-Ma
      significance given by the corresponding value of the legend
      pixel that is closest to the skymap pixel. The closeness is
      measured by a euclidean distance in the RGB space. The
      coordinates of the pixels were transformed successively by an
      inverse Mollweide projection, an equatorial to galactic
      coordinate transformation, and a Mollweide projection to create
      the new skymap shown in this figure. \label{fig}}
    \end{figure}

  The most recent search for hot spot
anisotropies is a joint effort by the Auger and TA collaborations
considering 840 events recorded by Auger with $E_{\rm Auger} >
10^{10.6}~{\rm GeV}$ and 130 events recorded by TA with $E_{\rm TA} >
10^{10.73}~{\rm GeV}$~\cite{Biteau:2019aaf}. Before proceeding, we
pause to note that even though the techniques for assigning energies to
events are nearly the same in both experiments, there are differences
as to how the primary energies are derived at Auger and TA, with
systematic uncertainties in the energy scale of the experiments
amounting to about $14\%$ and $21\%$ respectively, corresponding to
about $70\%$ uncertainty in the flux above a fixed energy
threshold. By comparison, the uncertainties on the respective
exposures are minor ($\lesssim 1\%$ and $\simeq 3\%$,
respectively). Therefore, it is necessary to cross-calibrate the
energy scales of the two datasets to avoid introducing a spurious
North/South asymmetry due to an energy scale mismatch. This is
accomplished by exploiting the wide declination band ($- 16^\circ \lesssim
\delta \lesssim +45^\circ$) where the two datasets overlap. Regardless of
the true arrival direction distribution, within a region of the sky
$\Delta \Omega$ fully contained in the field of view (FoV) of both
observatories, the sum over observed events $\sum_i 1/\omega({\bf n}_i
)$ (where $\omega$ is the directional exposure of each observatory in
the direction ${\bf n}_i$, in ${\rm km \, yr}$ units) is an unbiased
estimator of $\int_{\Delta \Omega} \Phi({\bf n}) \, d {\bf n}$ (where
$\Phi$ is the directional UHECR flux integrated above the considered
energy threshold, in ${\rm km^{-2} \, yr^{-1} \, sr^{-1}}$ units) and
should be the same for both experiments except for statistical
fluctuations. This criterium is generally adopted to cross-calibrate
the energy scales and to determine $E_{\rm Auger}$ and $E_{\rm TA}$
such that the Auger flux above $E_{\rm Auger}$ matches the TA flux
above $E_{\rm TA}$.\footnote{Actually, the region of the sky which is
  mostly used spans the declination band $-12^\circ \leq \delta \leq
  +42^\circ$. This is because including directions too close to the
  edge of the FoV of one of the observatories would result in larger
  statistical fluctuations due to very large values of
  $1/\omega(\bf{n}_i)$ near the edge.} The most significant excesses observed in a $20^\circ$
search are at: $(l, b)
\approx (303.0^\circ, 12.9^\circ)$ and ($l, b) \approx
(162.5^\circ,44.4^\circ)$, with local (Li-Ma~\cite{Li:1983fv}) statistical significance for
the rejection of the null (background only) hypothesis of $4.7\sigma$
and $4.2\sigma$, respectively. The Li-Ma significance
map of this data-sample is shown in 
Fig.~\ref{fig}.  The most significant hot spot is near the
location of starburst galaxies NGC 4945 and M83. The starburst galaxy M82 is at
the northern edge of the TA hot spot. A warm spot is also visible in
the skymap near the direction of the closest starburst NGC 253. In
closing, we note that the
clear and convincing evidence for the correlation between UHECRs
 and SBGs is further supported by a solid framework for
 particle acceleration to the highest observed
 energies~\cite{Anchordoqui:2018vji,Anchordoqui:2018qom,Anchordoqui:2019mfb}.
 In our calculations we will then assume that SBGs are {\it the} sources of UHECRs.

\section{Results and Conclusions}

The global structure of the GMF can be divided into the halo and disk
components. Each component can further be subdivided into a coherent
regular field $B_{\rm reg}$, which yields directional deflections and
a random field $B_{\rm rand}$. Pshirkov, Tinyakov, Kronberg and
Newton-McGee (PTKN) used data from the NRAO VLA Sky Survey rotation
measures catalog~\cite{Taylor:2009np} to constrain the
GMF~\cite{Pshirkov:2011um}. The observed distribution of RMs over the
sky disfavors ring disk models. A spiral disk and anti-symmetric halo
structure best fit the data. Targeted observations of Galactic
structures and high resolution synchrotron mapping of external
galaxies such as in the CHANGES survey~\cite{Irwin:2012by} as well as
sky maps of polarized and unpolarized Galactic synchrotron emission
from WMAP were considered by Jansson and Farrar (JF) to complement RMs
and develop a sophisticated GMF
model~\cite{Jansson:2012pc,Jansson:2012rt}. (The 7-year WMAP
  synchrotron maps~\cite{Gold:2010fm} were used in the original JF
  analysis~\cite{Jansson:2012pc,Jansson:2012rt}; the 9-year final WMAP
  data release~\cite{Bennett:2012zja} and the Planck 2015 data
  release~\cite{Adam:2015wua} were considered in the update
  of~\cite{Unger:2017wyw}.) The JF model contains three distinct
components: {\it (1)}~a coherent large-scale field, with disk, halo
and out-of-plane components, {\it (2)}~a fully random field specified
by its spatially-varying rms field strength, and {\it (3)}~a
``striated'' random field. {\it (1.a)}~The disk component of the
coherent field is toroidal in the inner ``molecular ring'' region from
$3-5~{\rm kpc}$, beyond which it has a logarithmic-spiral
geometry. The typical strength of the coherent disk field in the
magnetic arms is roughly $1~\mu{\rm G}$, with maximum values of a few
$\mu {\rm G}$.  {\it (1.b)}~The regular halo component is modelled as
oppositely directed coherent toroidal fields above and below the
Galactic Plane. The sense of rotation below the plane (Southern
hemisphere) is in the same direction as the rotation of the disk. The
toroidal fields reach their maximum strength ${\cal O} (\mu {\rm G})$
about 1~kpc away from the
plane, beyond which they decline slowly reaching half their peak value
at about 5~kpc. {\it (1.c)}~The out-of-plane halo component of the
coherent field or (X-field) is azimuthally symmetric and poloidal; its
strength is $5~\mu{\rm G}$ at the Galactic center, diminishing rather
slowly with distance from the Galactic plane. The radial-scale length
of the X-field is about ${\rm 3kpc}$ and its value in the solar
neighborhood is approximately $0.2~\mu{\rm G}$. The sense of the halo
toroidal fields are consistent with their resulting from differential
rotation of the coherent poloidal X-field.  The disk, toroidal halo, and X fields were
required to be separately divergenceless, so their free parameters
could be adjusted independently. {\it (2)}~The random field
is modelled as a superposition of a disk component whose spiral-arms
are the same as those adopted for the coherent field, but with
independently fitted rms field strength, and a smooth halo
component. The  halo field has an azimuthal component,
which can be characterized by its overall rms strength, and radial and
vertical scale lengths. The understanding of the random field
structure and its maximum strength are muddled by the uncertainty in
$n_{\rm CRe}$. The best-estimate for the maximum field strength is
${\cal O}(10~\mu{\rm G})$. {\it (3)}~The striated
field is aligned with the local coherent field and its rms strength is
locally proportional to the coherent field strength, that is
$B_{\rm stri}^2 \propto B_{\rm reg}^2$.

Using (\ref{delta}), the expected deflections of UHECRs when crossing a distance $L$ of the Galaxy are estimated to be~\cite{Mollerach:2017idb}
\begin{equation}
\delta \theta_{\rm G} \simeq 10^\circ \ Z \left(\frac{E}{10^{10}~{\rm
        GeV}} \right)^{-1} \left| \int_0^L \frac{d\vec x}{\rm kpc}
    \times \frac{\vec B}{2~\mu{\rm G}} \right| \, .
\end{equation}
This implies that particles in the energy range
$10^{10.6} \lesssim E/{\rm GeV} \lesssim 10^{11.3}$, which would
suffer deflections of $\sim 13^\circ$, are most likely CNO, with
$Z\leq 8$. Note that the helium contribution to the flux will be largely
suppressed because of energy loss during
propagation~\cite{Soriano:2018lly}.  In our calculations we then take
as fiducial a particle rigidity of $10^{10}~{\rm GV}$.  For,  $Z=1$
and $E = 10~{\rm EeV}$ we have $c_2 = 3 \times 10^{-23}{\rm
  rad}/(\mu{\rm G} \, {\rm cm})$~\cite{Pshirkov:2013wka}. For further reference,  $c_1/c_2 =
0.9~{\rm cm}$.   In
Fig.~\ref{fig} we show the expected deflection for protons of
$E = 10^{10}~{\rm GeV}$ according to the PTKN~\cite{Pshirkov:2011um} and JF~\cite{Jansson:2012pc,Jansson:2012rt} GMF
models~\cite{Mollerach:2017idb,Farrar:2017lhm,Anjos:2018mgr}. It is
clear from the figure that the expected deflections are consistent (at
least at the qualitative level) with the observed excess in Auger data.  However, on the assumption that the TA hot spot originates in
the SBG M82 we conclude that the expected deflections exhibit a poor
representation of the TA data. More concretely, the expected
deflections of UHECRs entering the Galaxy from the direction of M82
shown in Fig.~\ref{fig} are towards the Galactic north-east of the
M82, whereas the center of the TA hot spot is in the Galactic
north-west direction of the source. Note that this is the case for
both GMF models shown in the figure, and also for all possible
variations of the JF configuration discussed in~\cite{Unger:2017wyw}.
We conclude that if the starburst hypothesis is validated by future
Auger data, then M82 must be a powerful source of UHECRs and must
dominate the contribution to the TA hot spot. This can be used to
constrain the GMF with future UHECR data, for which the nuclear
composition of each event is known.

The Probe Of Extreme Multi-Messenger Astrophysics (POEMMA) is a NASA
space-based mission~\cite{Olinto:2017xbi}. POEMMA is optimized for the measurement of
extensive air showers (EASs) from UHECRs using the stereo air
fluorescence technique, and from neutrino induced upward-going EASs
via optical Cherenkov detection. POEMMA makes
observations in umbra and in low moonlight conditions. POEMMA is designed to reach
unprecedented geometrical apertures $> 10^6~{\rm km^2 \, sr} 
  \; {\rm yr}$,
which after duty cycle corrections, correspond to annual exposures of
more than $10^5~{\rm km \, sr \, yr}$ at the highest energies. POEMMA
is composed of two identical satellites flying in formation with the
ability to observe overlapping regions during moonless nights at
angles ranging from nadir to just above the limb of the Earth.  The
satellites will fly at an altitude of about 525~km, with separations
ranging from 300~km for stereo fluorescence UHECR observations to
25~km when pointing at the Earth's limb for both fluorescence and
Cherenkov observations of UHECRs and cosmic neutrinos. The satellites
will orbit the Earth with a period of 95 minutes, orbiting the Earth
$\sim 15$~times per day. POEMMA has full-sky coverage due to its orbit at
525~km altitude and $28.5^\circ$ inclination and the very large
field-of-view ($45^\circ$) for each satellite. The ability of the
space-based POEMMA telescopes to tilt towards the northern or southern
hemisphere allows for the sky exposure can be enhanced for a specific
hemisphere. Likewise, it is easy for POEMMA to view north or south for
a sequence of orbital periods to further tailor the UHECR sky coverage
for possible source locations.

The atmospheric column depth at which the longitudinal development of
an EAS reaches maximum, $X_{\rm max}$, is a powerful observable to
determine the UHECR nuclear composition. Detailed simulations of POEMMA's UHECR exposure, angular resolution,
and $X_{\rm max}$ resolution were performed using the instrument
design~\cite{Anchordoqui:2019omw}. POEMMA stereo observations of  EASs will
have high angular resolution $\lesssim 1^\circ$ for
$E > 10^{10.5}~{\rm GeV}$. The fine angular resolution leads to accurate
3-dimensional reconstruction of the EASs, with
energy resolution of $\sim 20\%$ and $X_{\rm max}$ resolution of $\sim
35
(10^{10.6}~{\rm GeV}/E)^{0.55}~{\rm g/cm^2}$. The
event-by-event composition measurements together with the full-sky
distribution of arrival directions will provide a profitable
data-sample for constraining GMF models.

In summary, the GMF has always been seen as a hindrance for
charged particle astronomy. In the spirit of~\cite{Anchordoqui:2001bs},  here we have shown that now we can turn
things around and use UHECR deflections to constrain GMF models.

\section*{Acknowledgments}
We thank our colleagues of the POEMMA and Pierre Auger collaborations
for discussion.  This work has been supported by the U.S. National Science Foundation
(NSF Grant PHY-1620661), the National Aeronautics and Space
Administration (NASA Grant 80NSSC18K0464), as well as by grants PGC2018-095512-B-I00, AYA2017- 92402-EXP, iLink 2017-1238, and SGR 2017-1383.

\end{document}